\documentclass[sn-mathphys-num]{sn-jnl}

\usepackage{graphicx}%
\usepackage{subcaption}
\usepackage{multirow}%
\usepackage{amsmath,amssymb,amsfonts}%
\usepackage{amsthm}%
\usepackage{mathrsfs}%
\usepackage[title]{appendix}%
\usepackage{xcolor}%
\usepackage{textcomp}%
\usepackage{manyfoot}%
\usepackage{booktabs}%
\usepackage{algorithm}%
\usepackage{algorithmicx}%
\usepackage{algpseudocode}%
\usepackage{listings}%
\usepackage{multirow}

\theoremstyle{thmstyleone}%
\theoremstyle{thmstyletwo}%

\theoremstyle{thmstylethree}%

\begin{document}

\title{Increasing the density limit with ECRH-assisted Ohmic start-up on EAST}

\author[1]{\fnm{Jiaxing} \sur{Liu}}

\author*[1,2]{\fnm{Ping} \sur{Zhu}}\email{zhup@hust.edu.cn}

\author[3]{\fnm{Dominique Franck} \sur{Escande}}

\author[4]{\fnm{Wenbin} \sur{Liu}}
\author[1]{\fnm{Shiwei} \sur{Xue}}
\author[4]{\fnm{Xin} \sur{Lin}}
\author[4]{\fnm{{Panjun}} \sur{{Tang}}}
\author[4]{\fnm{Liang} \sur{Wang}} 
\author*[4]{\fnm{Ning} \sur{Yan}} \email{yanning@ipp.ac.cn}
\author[4]{\fnm{{Jinju}} \sur{{Yang}}}
\author[4]{\fnm{{Yanmin}} \sur{{Duan}}}
\author[4]{\fnm{{Kai}} \sur{{Jia}}}
\author[4]{\fnm{{Zhenwei}} \sur{{Wu}}}
\author[4]{\fnm{{Yunxin}} \sur{{Cheng}}}
\author[4]{\fnm{Ling} \sur{Zhang}}
\author[4]{\fnm{Jinping} \sur{Qian}}
\author[4]{\fnm{Rui} \sur{Ding}}
\author[4]{\fnm{Ruijie} \sur{Zhou}}
\author[4]{\fnm{and the EAST} \sur{team}}
\affil[1]{\orgdiv{State Key Laboratory of Advanced Electromagnetic Technology,
International Joint Research Laboratory of Magnetic Confinement Fusion
and Plasma Physics, School of Electrical and Electronic Engineering}, \orgname{Huazhong University of Science and Technology}, \orgaddress{
        \city{Wuhan}, \postcode{453500}, \state{Hubei}, \country{China}}}

\affil[2]{\orgdiv{Department of Nuclear Engineering and Engineering Physics}, \orgname{University of Wisconsin-Madison}, \orgaddress{
        \city{Madison}, \postcode{53706}, \state{Wisconsin}, \country{United States of America}}}

\affil[3]{
    \orgname{Aix-Marseille Universit$\acute{\text{e}}$}, \orgaddress{\street{{CNRS, PIIM, UMR 7345}}, 
\state{Marseille}, \country{France}}}

\affil[4]{\orgdiv{Institute of Plasma Physics},
    \orgname{Chinese Academy of Sciences}, \orgaddress{
        \city{Hefei},
        \postcode{230031}, \state{Anhui}, \country{China}}}


\abstract{High plasma density operation is crucial for a tokamak
    to achieve energy breakeven and a burning plasma. However,
    there is often an empirical upper limit of electron density in tokamak operation, namely the Greenwald density limit $n_{\text{\tiny G}}$, above which tokamaks generally disrupt. Achieving high-density operations above the density limit has been a long-standing challenge in magnetic confinement fusion research.
    Here, we report experimental results on EAST tokamak
    achieving the line-averaged electron density in the range of 1.3 $n_{\text{\tiny G}}$ to 1.65 $n_{\text{\tiny G}}$,while the usual range in EAST is (0.8-1.0)$n_{\text{\tiny G}}$. This is performed with ECRH-assisted
    Ohmic start-up and a sufficiently high initial neutral density. This is motivated by and consistent with predictions of a recent plasma-wall self-organization (PWSO) theory, that increasing ECRH power or
    pre-filled gas pressure leads to lower plasma
    temperatures around divertor target and higher density limits.
    In addition, the experiments are shown to operate in the density-free
    regime predicted by the PWSO model. These results suggest a promising scheme for substantially increasing
    the density limit in tokamaks, a critical advancement
    toward achieving the burning plasma.}




\maketitle

\section{Introduction}\label{sec1}

Achieving high plasma density is essential for satisfying the
Lawson criterion in magnetic confinement fusion, yet density
limits often constrain tokamak performance, potentially leading
to disruptive terminations \cite{Greenwald2002}. Over decades,
understanding and predicting these limits have remained a
focal point of fusion research. For past decades, the Greenwald
density limit $\displaystyle
    n_{\text{\tiny G}}\left(\text{m}^{-3}\right)={I_\text{p}\left(\text{MA}\right)\times 10^{20}}/{\pi
    \left(a\left(\text{m}\right)^2\right)}$ has been widely used as an empirical
scaling law for the tokamak density limit {(DL)} \cite{Greenwald2002}.
However, this scaling law does not account for the
heating power dependence of the density limit that has been observed in
many experiments \cite{Borrass1991,Staebler1993ComparisonOD,Mertens1997,Rapp1999,huber2017,manz2023} and confirmed by many
theoretical models \cite{Zanca2017,Zanca2019,Zanca2022,Giacomin2022,Singh2022,Stroth2022,Diamond2023}.
For example, the power balance model~{\cite{Zanca2017,Zanca2019,Zanca2022}} considering impurity radiation
effects introduces a modified scaling for the density limit,
$(I_\text{p}P/a^4 )^{(4/9)}$, where  $P$ is the heating power.
This radiative scaling is in better agreement with the tokamak
experimental databases compared with the Greenwald density limit (Fig.~4 of \cite{Zanca2022}). Besides, more experiments have exceeded the Greenwald density limit, such as those on T-10~\cite{Esipchuk2003}, FTU~\cite{,Pucella2013}, D\uppercase\expandafter{\romannumeral3}-D~\cite{ding2024}, and MST~\cite{Hurst2024}.

{In the wake of \cite{Zanca2017,Zanca2019,Zanca2022}, a} recent plasma-wall self-organization (PWSO) theory was built under the assumption that the primary factors influencing the power balance limits stem from impurity radiation, which is largely controlled by
plasma-wall interactions \cite{escande2022}.
This theory provides new insights into the interaction between
plasma dynamics and wall conditions through
impurity radiation \cite{escande2022}. It shows that there are two basins of attraction for the density: one with a density limit and one without such a limit. In particular, the power dependent scalings of density limit has been derived from this theory and are able to match the experimental results in multiple tokamak devices~\cite{jxliu2025a}.  
{Besides, a relatively higher density
limit is reached in stellarator when the start-up is performed
by using higher ECRH power \cite{Klinger2019,Wolf2019}. This higher density limit
might be due to their mode of breakdown at start-up phase: the
massive use of ECRH power with high neutral density producing less impurities \cite{Zanca2017,escande2022}.} Both PWSO theory and stellarator results were an incentive to perform experiments in J-TEXT, which directly validated key aspects
of PWSO theory in electron cyclotron resonance heating
(ECRH)-assisted Ohmic start-up discharges {\cite{jxliu2023,yhding2024}}.
These experiments
demonstrate that increasing ECRH power or pre-filled gas
pressure enhances the achievable density limit by
decreasing impurity radiation and increasing target region
plasma temperatures but by remaining in the density limit basin of PWSO theory.

Building on the success of these experiments, the EAST tokamak
provides a unique platform to extend the validation of the PWSO
model. With its advanced capabilities, especially the tungsten
plasma-facing components, the EAST tokamak enables the exploration of density
limits under conditions distinct from those of J-TEXT and more adequate to check PWSO, since physical sputtering dominates over the chemical one, which might enable reaching the basin without density limit.
This paper presents EAST experimental results,
with a focus on the role of ECRH power and pre-filled gas pressure
in achieving high-density regimes. Comparative analysis with
the PWSO theory is conducted to elucidate the mechanisms
underpinning observations, and to further refine our
understanding of density limits in tokamaks. The EAST experimental results show that increasing the pre-filled neutral gas pressure and/or the ECRH power can increase the density limit up to $1.3n_{\text{\tiny G}}$ to $1.65n_{\text{\tiny G}}$, while the usual range in EAST is (0.8-1.0)$n_{\text{\tiny G}}$~\cite{xwzheng2013,jilei2024}. The experimental data for density limit $n_c$ and plasma temperature around divertor $T_t$ are consistent quantitatively with the prediction of PWSO theory. Moreover, as expected, these discharges are found to operate in the density-free regime of PWSO theory, which implies that a substantial enhancement in the density limit of tokamaks may be attainable through achieving detachment without actively introducing impurities.

The remainder of this paper is organized as follows:
section 2 introduces the experimental set-up and results. Section 3 presents the comparison between experimental results and predictions of PWSO model. Section 4 offers several discussion points and implications of our findings for fusion research. Section 5 concludes with a summary and discussion.
\section{Experimental results}\label{sec2}

The experiments in this work are conducted in a lower single null divertor configuration
with purely or ECRH-assisted Ohmic start-up. The details of ECRH-assisted Ohmic start-up can be found in reference \cite{runze2024,wenbin2024}. The plasma density is feedback-controlled using gas puffing~\cite{xwzheng2013} and the  \#2 ECRH system is used for auxiliary heating~\cite{wyxu2024}. The wall conditioning technique, lithiation~\cite{jshu2014}, is used in our experiments. 
The typical parameters in EAST experiments are as follows: plasma current $I_p\sim250$~kA, plasma elongation ratio $\kappa\sim1.5$, triangularity $\delta\sim0.5$ and toroidal magnetic field $B_0\sim2.5$~T.

\begin{figure}[htbp]
    \centering
    \includegraphics[width=1.05\linewidth]{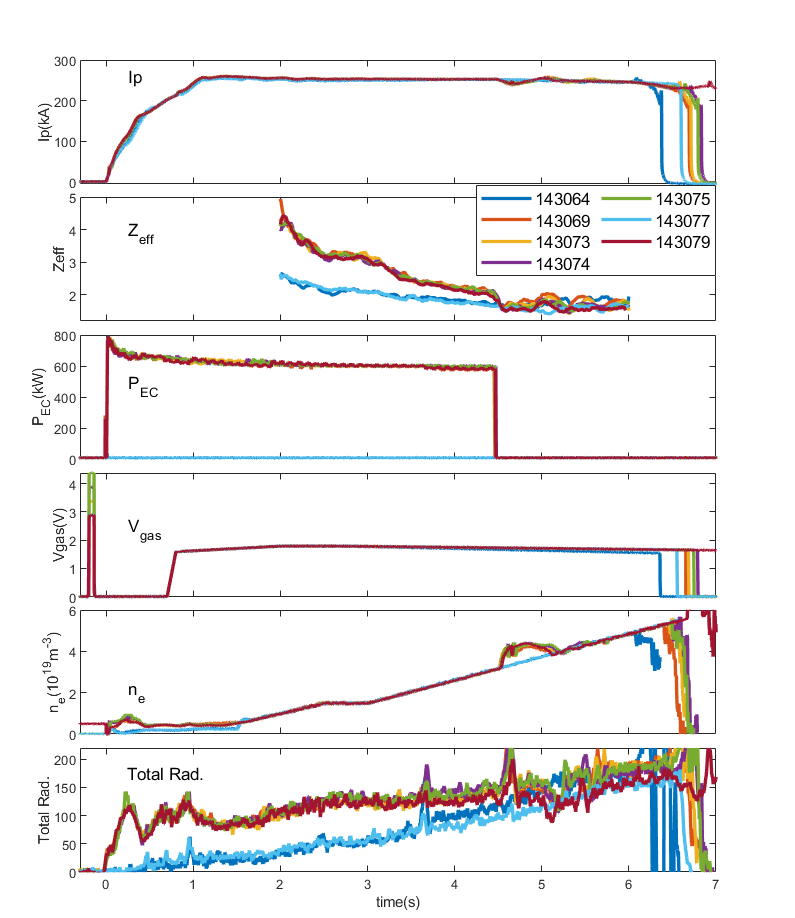}
    \caption{Time histories of key parameters in density limit discharges with varied ECRH power. Here
        $I_p$: plasma current;
        Zeff: averaged effective charge number;
        $P_\text{EC}$: injected ECRH power;
        $V_\text{gas}$: voltage applied to control gas puffing, which is positively correlated with the  puffed gas amount;
        $n_{e}$: line-averaged electron density measured using the central chord of HCN interferometer;
        Total Rad.: relative intensity of total radiation measured using bolometer.
    }
    \label{fig:discharges_differentPEC_121224}
\end{figure}

There are two main series of discharges in our experiments {(Table~\ref{tab1})}, {namely, the} one with varied pre-filled neutral gas pressure at a fixed ECRH power and the other with varied ECRH heating powers at a fixed pre-filled pressure. We selected a reference discharge \#143069 to serve as a point of comparison, {which {is based on an} ECRH-assisted Ohmic start-up scenario {similar to} those reported in references \cite{runze2024,wenbin2024}}. The time history of its key parameters, including total radiation levels, plasma current $I_p$, ECRH power and line-available electron density $n_e$, are shown in Fig.~\ref{fig:discharges_differentPEC_121224}. The line-averaged density, $n_e$, used in this article is measured by the central chord of the HCN interferometer unless otherwise specified. This discharge starts at 0.0 s, with a toroidal magnetic field of 2.5 T, an initial gas puffing voltage of about 3 V with the corresponding number of injected deuterium (D$_2$) of $6.6\times10^{19}$, an ECRH power of about 600 kW with the pulse width [0.0 s, 5.0 s], and a plasma current of 250 kA. Hydrogen gas is injected as the working gas, and no other kind of gas or impurity is injected throughout the discharge. During the current plateau period, the gas injection continues until the plasma density limit of about 1.5~$n_{\text{\tiny G}}$ is reached. At this point, the plasma density starts to decrease and then rapidly drops to zero, without any prior manifestation of MHD activities. During the whole discharge, the feedback control technique is used to make sure the time evolution of line-averaged electron density follow the designed path~\cite{Yuan2010}.

\begin{figure}[htbp]
    \centering
    \begin{subfigure}{0.49\linewidth}
        \centering
        \includegraphics[width=\linewidth]{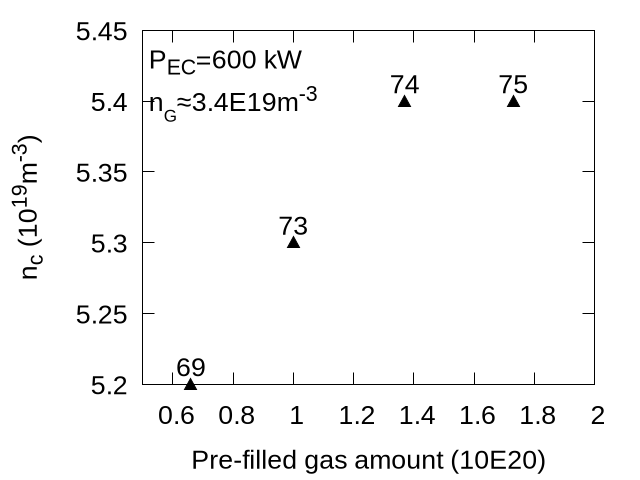}
        \caption{}
        \label{fig:nc_VS_gas-pressure_different-pre-filled-gas_121224}
    \end{subfigure}
    \begin{subfigure}{0.49\linewidth}
        \centering
        \includegraphics[width=\linewidth]{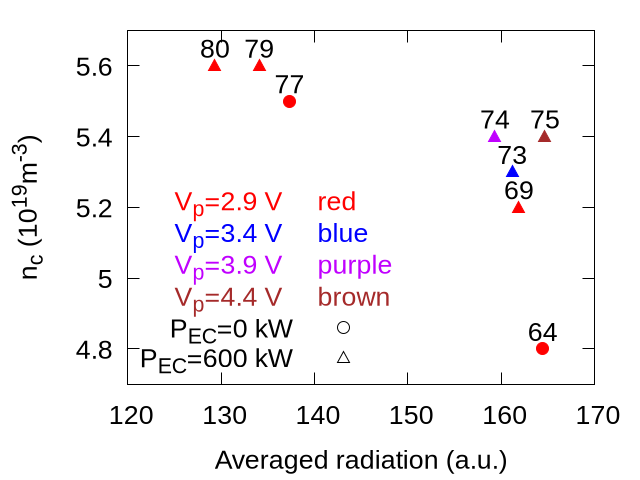}
        \caption{}
        \label{fig:ne_bolo-rad_022725}
    \end{subfigure}
    \begin{subfigure}{0.49\linewidth}
        \centering
        \includegraphics[width=\linewidth]{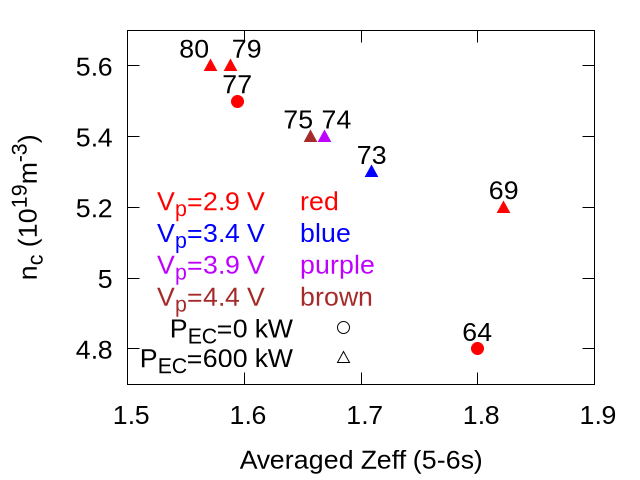}
        \caption{}
        \label{fig:ne_Zeff_022725}
    \end{subfigure}
    \begin{subfigure}{0.49\linewidth}
        \centering
        \includegraphics[width=\linewidth]{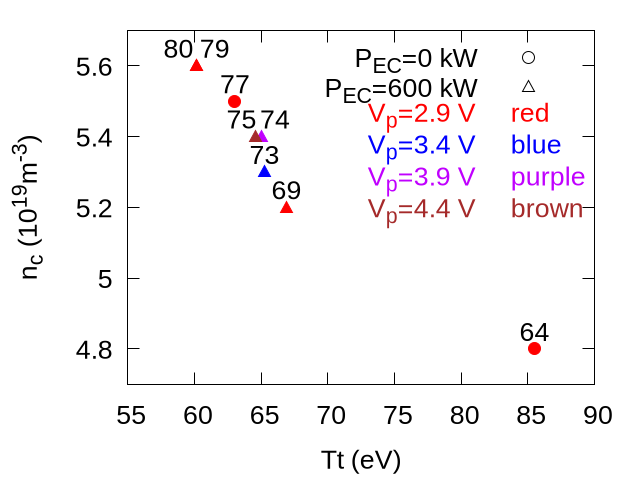}
        \caption{}
        \label{fig:ne_Tdiv_022725}
    \end{subfigure}
    \caption{The density limit {$n_\text{c}$} and {(a) the corresponding particle number of pre-filled gas {(D$_2$)} with an ECRH power of about 600 kW; (b) the averaged relative radiation in time interval (5s-6s); (c) the averaged effective charge number $Z_\text{eff}$; (d) the averaged plasma temperature around the lower outer divertor $T_\text{t}$ with various pre-filled gas pressure levels in the start-up phase and 2 different ECRH powers}. The number label next to each symbol denotes the last 2 digits of the shot number 1430xx.}
    \label{fig:4figures}
\end{figure}

For discharges with varied pre-filled neutral gas and ECRH power of about 600 kW, the experimental results indicate that the density limit increases with the pre-filled gas pressure until saturation (Fig.~\ref{fig:nc_VS_gas-pressure_different-pre-filled-gas_121224}). This saturation should be attributed to the limited ECRH power, instead of an absolute limit. Higher ECRH power should lead to higher pressure thresholds where saturation occurs, much like the relation between gas pressure threshold and ECRH power during the startup phase~\cite{runze2024,wenbin2024}. This enhancement of density limit is related to the decrease of total radiation power (Fig.~\ref{fig:ne_bolo-rad_022725}) and the higher cleanliness of plasma which is indicated by the effective charge number $Z_\text{eff}$ (Fig.~\ref{fig:ne_Zeff_022725}). The effective charge number and total radiation power $Z_\text{eff}$ are averaged in an interval [5s-6s] when the plasma is close to the disruption. The cleanliness of plasma and the enhanced density limit should be mainly influenced by the plasma-wall interaction, which is related to the plasma temperature around the divertor target. This can be inferred from Fig.~\ref{fig:ne_Tdiv_022725}, which shows the plasma temperature around the low-outer divertor target is lower when the pre-filled gas pressure is higher and the final density limit is higher.

\begin{figure}[htbp]
    \centering
    \includegraphics[width=\linewidth]{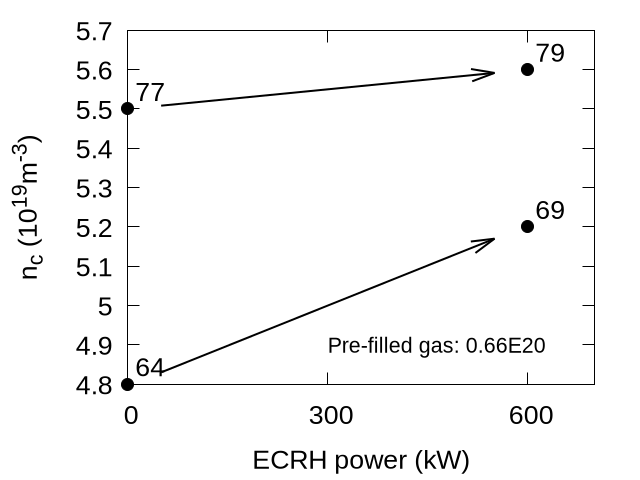}
    \caption{The density limit and the corresponding ECRH power
        for discharges with {pre-filled gas level of $0.66\times10^{20}$ deuterium molecules}. The number label next to each symbol denotes the last 2 digits of the shot number 1430xx.}
    \label{fig:nc_VS_PEC_different-PEC_121224}
\end{figure}

For discharges with varied ECRH heating powers and the lowest pre-filled gas pressure of all cases, the experimental results indicate that the density limit weakly increases with the ECRH power. For instance, from shot \#143064 to shot \#143069, density limit increases from $4.9\times 10^{19}\mathrm{m}^{-3}$ to $5.2\times 10^{19}\mathrm{m}^{-3}$ as the ECRH power increases from 0 to 600 kW; from shot \#143077 to shot \#143079, density limit increases from {$5.5\times 10^{19}\mathrm{m}^{-3}$} to {$5.6\times 10^{19}\mathrm{m}^{-3}$} with the ECRH power increasing from 0 to 600 kW (Fig.~\ref{fig:nc_VS_PEC_different-PEC_121224}). {The lowest pre-filled gas pressure might be the reason for the weak DL increase, since there {would be less amount of gas from the outset available for later density increase. Lower pre-filled gas pressure is also known to lead to lower accessible density regime with increasing ECRH power during start-up \cite{wenbin2024}}.}
This density limit enhancement is also related to the decrease of total radiation (Fig.~\ref{fig:ne_bolo-rad_022725}), the higher plasma cleanliness (Fig.~\ref{fig:ne_Zeff_022725}){, and the lower plasma temperature around the low-outer divertor target (Fig.~\ref{fig:ne_Tdiv_022725})}, which can all be attributed to the reduced strength of plasma-wall interaction.

\begin{figure}[htbp]
    \centering
    \begin{subfigure}{1\linewidth}
        \centering
        \includegraphics[width=\linewidth]{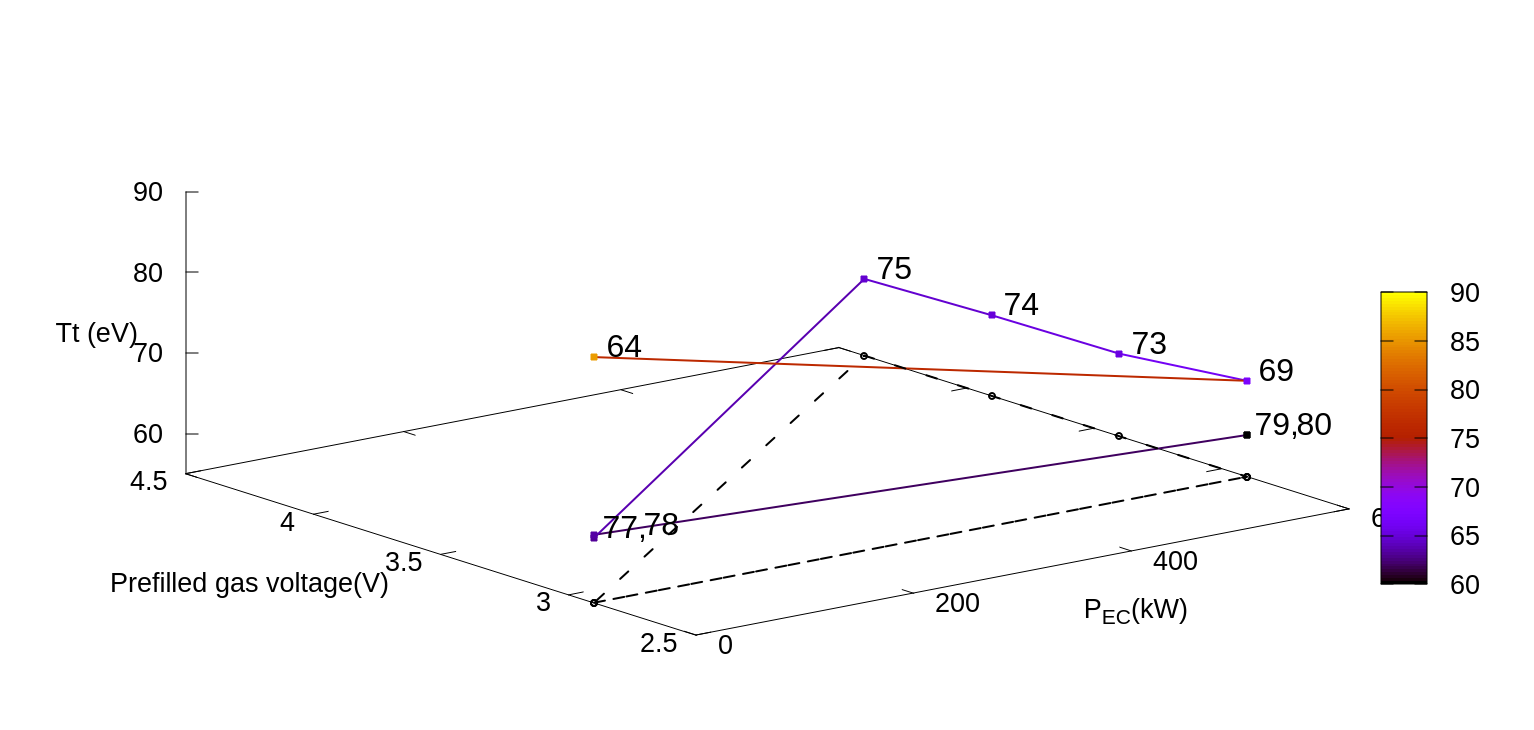}
        \caption{}
        \label{fig:Tt_3D_022725}
    \end{subfigure}
    \begin{subfigure}{1\linewidth}
        \centering
        \includegraphics[width=\linewidth]{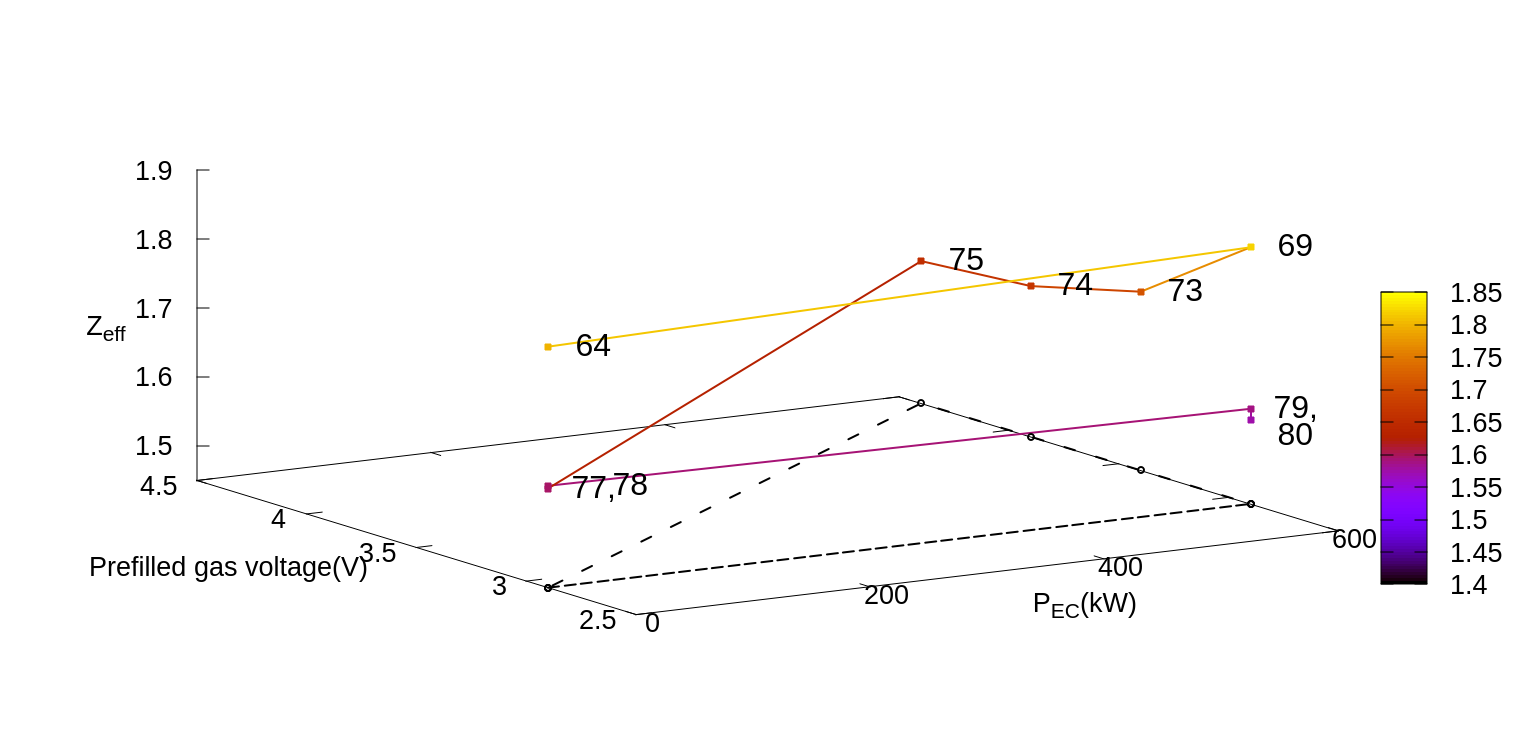}
        \caption{}
        \label{fig:Zeff_3D_022725}
    \end{subfigure}
    \begin{subfigure}{1\linewidth}
        \centering
        \includegraphics[width=\linewidth]{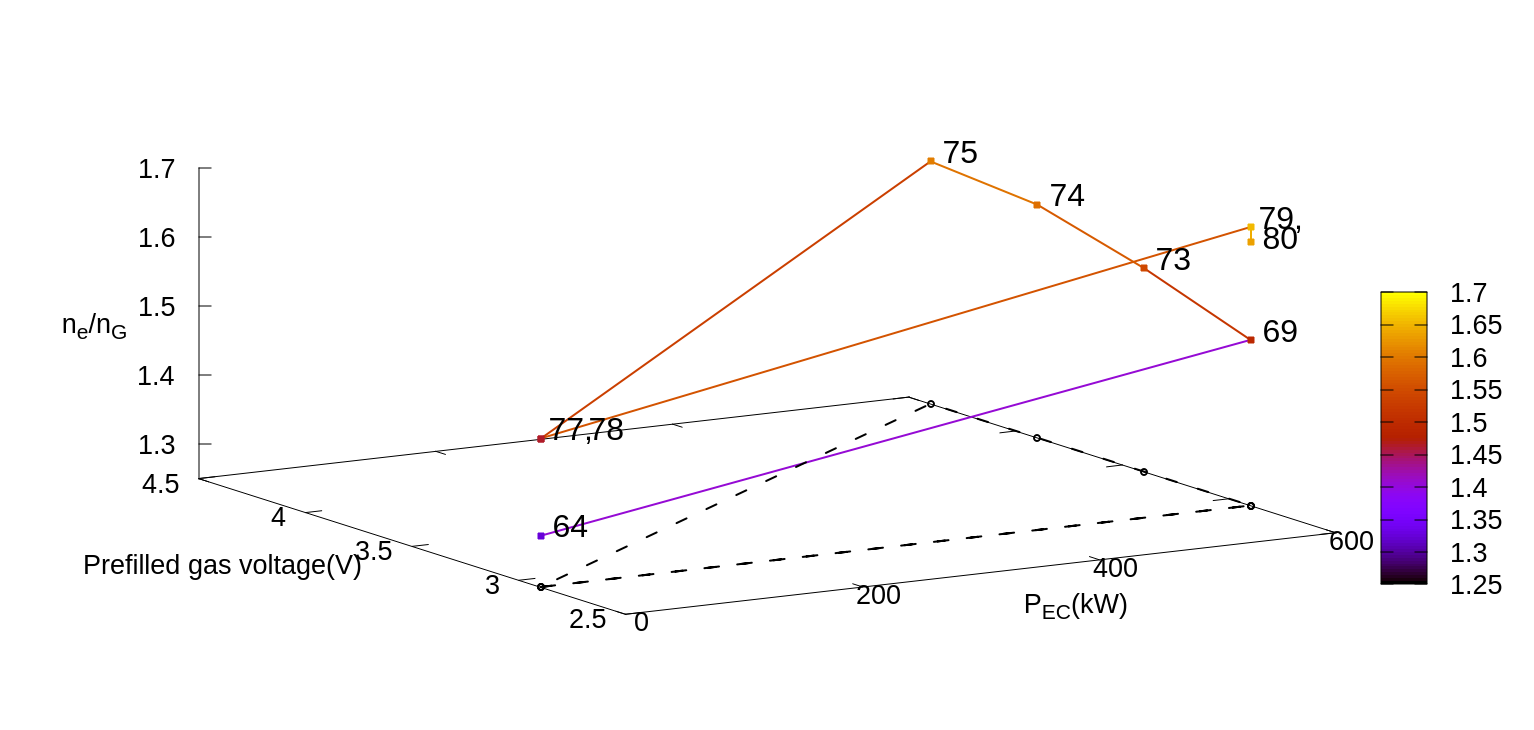}
        \caption{}
        \label{fig:ne_nG_3D_022725}
    \end{subfigure}
    \caption{3D diagram displaying {(a) the target region plasma temperature $T_t$, (b) the effective number $Z_\text{eff}$, and (c) the density limit $n_e/n_{\text{\tiny G}}$} of successive effective discharges as a function of both ECRH power $P_\text{EC}$ and prefilled gas voltage. The black dotted line is the projection of the colored line in the $P_\text{EC}$-pre-filled gas voltage plane. The number label next to each symbol denotes the last 2 digits of the shot number 1430xx. The color bar represents the value of $T_t$.}
    \label{fig:3figures}
\end{figure}

It is noted that shots \#143079 and \#143069 have identical input parameters, yet yield different plasma temperature around the divertor target or density limit. A similar observation can be made regarding shots \#143064 and \#143077. These differences may be due to alterations in wall conditions over time~\cite{Asakura2004}, because there are several density limit discharges between shots \#143064 and \#143079, which can have potential impacts on the wall conditions. The successive effective discharges shown in {Fig.~\ref{fig:Tt_3D_022725}}, Fig.~\ref{fig:Zeff_3D_022725}, and Fig.~\ref{fig:ne_nG_3D_022725} demonstrate a striking point that the target region plasma temperature $T_t$ and the effective charge number $Z_\text{eff}$ decreases with discharge number, and the density limit $n_e$ increases with discharge number. This indicates the wall condition is improving with time. Therefore, the above technique improves the DL, but this improvement is not a simple function of ECRH power and pre-filled gas pressure.

In summary, these experimental results demonstrate that increasing pre-filled gas or/and ECRH power leads to lower plasma temperature around the divertor target and higher density limit well above the Greenwald density limit (Fig.~\ref{fig:ne_Tdiv_022725}). This enhancement of density limit is related to the improved wall condition.
In the following, we will compare the experimental results with the predictions of plasma-wall self-organization theory.

\section{Comparison with PWSO theory}\label{sec3}
PWSO theory describes the plasma-wall interaction through the relationship between sputtered impurities' radiation and heating power \cite{escande2022,jxliu2023}.
\subsection{PWSO 0D model and comparison}
The basic idea of the PWSO theory ({section 4.1 of \cite{escande2022}}) is that
the existence of a time delay in the feedback loop relating impurity radiation
and impurity production on divertor/limiter plates yields the following equation in a 0D model
\begin{equation}
    \label{equ:0Diter}
    R_+ =\alpha(P-R)
\end{equation}
where $P$ is the total input power
to plasma, $R$ the total radiated power, and $R_+$
the delayed radiation power during the next
cycle of the feedback loop.
The coefficient $\alpha$ quantifies the radiation power $R_+$
generated by the impurity produced from the plasma-wall interaction that is
proportional to the deposition of the outflow power $\left( P-R \right)$ onto
the wall targets, which can be modeled as \cite{escande2022,jxliu2023}
\begin{equation}
    \label{eq:alpha}
    \alpha=\frac{f \lambda}{a D_{\perp} T_\text{t}} I\left(T_\text{t}\right) \int_{0}^{a} r n(r) R_\text{coe} \mathrm{d} r
\end{equation}
Here $a$ is the plasma minor radius, $n$ is the electron density, $D_\perp$ is the perpendicular
diffusion coefficient, $T_\text{t}$ is the plasma temperature around the
target plate location, $f$ is the fraction of the sputtered atoms that reach the main plasma
and become ionized at a distance $\lambda$  inwards from the
plasma target location, $R_\text{coe}$ is the impurity radiation rate coefficient, and $I\left(T_\text{t}\right)$
is an average of the yield function of impurity $Y\left(E\right)$ over the
energies of the impinging particles
\begin{equation}
    I\left(T_\text{t}\right)=\sqrt{\frac{m}{2\pi T_\text{t}}} \int_{0}^{\infty} Y\left(\frac{m v^{2}}{2}+\gamma T_\text{t}\right) \exp \frac{-m v^{2}}{2T_\text{t}} \mathrm{~d} v
    \label{eq:ITt}
\end{equation}
where  $\gamma$ is the total energy transmission coefficient
\cite{Stangeby2000}, $\gamma T_t$ is a measure of the Debye shield length,
and $m$ is the ion mass.
The fixed point of Eq.~$\left(\ref{equ:0Diter}\right)$ $R= R_+$
corresponds to the plasma-wall self-organization equilibrium, which becomes unstable for $\alpha>1$ as
predicted from Eq.~$\left(\ref{equ:0Diter}\right)$. So the the
threshold $\alpha=1$ establishes an upper radiation density limit
\begin{equation}
    n_{c}=\frac{2 D_{\perp}}{f \lambda \operatorname{Rad}[T]} \frac{T_\text{t}}{I\left(T_\text{t}\right) a}
    \label{equ:0dnlim}
\end{equation}
that can be reached for a
ratio of total radiated power to total input power as low as 1/2
\cite{escande2022}.
There are two density limit basins of PWSO. One is the regime of
density limit corresponding to the higher temperature of target, whereas
the other is the regime of density freedom corresponding to the lower
temperature of target, in particular in machines where the target plates
are made of high-Z materials~\cite{escande2022,jxliu2023}.
{For high-discharge-number impurity, such as tungsten,  the yield function $Y(E)$ is dominated by the physical sputtering {\cite{qzhang2022}}. So in the following calculation, the contribution from chemical sputtering is ignored. The interpolating functions for $Y(E)$ at normal incidence are provided in the Eq.~$\left(15\right)$ of \cite{YAMAMURA1996} for physical sputtering
  \begin{equation}
    \label{eq:yphy}
    Y_\text{phy}(E)=0.042 \frac{Q\left(Z_{2}\right) \alpha^{*}\left(M_{2} /
        M_{1}\right)}{U_{\text{s}}}\frac{S_{\text{n}}(E)}{1+\Gamma k_{\text{e}} \epsilon^{0.3}}
    \times{\left[1-\sqrt{\frac{E_{\text{th}}}{E}}\right]^{s} }
  \end{equation}
    Here the numerical coefficient $0.042$ is in unit of \r
      A$^{-2}$, $Z_1$ and $Z_2$ are the atomic numbers, $M_1$ and $M_2$ are the
    masses of the projectile and the target atoms, respectively, $S_\text{n}$ is
    the reduced nuclear stopping cross section, $U_\text{s}$ is the surface
    binding energy of the target solid, $k_\text{e}$ is the Lindhard electronic
    stopping coefficient, $E$ is the projectile energy, $E_\text{th}$ is the
    threshold energy for sputtering, $\epsilon$ is the reduced energy
    {${a_L}E{M_2}/\left({(M_1+M_2)}{Z_1Z_2e^2}\right)$}, the $\Gamma$ factor  has the
    form $W\left(Z_2 \right)/\left(1+\left( M_1/7\right)^3 \right)$, and $W$
    and $Q$ are dimensionless fitting coefficients. }

\begin{figure}[htbp]
    \centering
    \includegraphics[width=\linewidth]{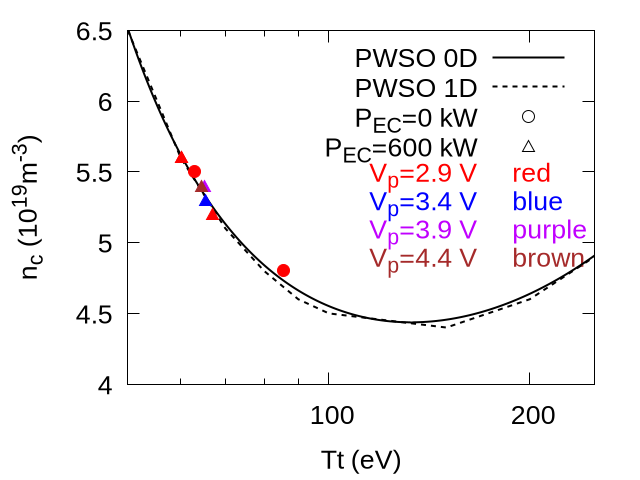}
    \caption{The density limits as functions of the target region plasma temperature $T_\text{t}$ predicted from the PWSO 0D (solid line) and 1D (dashed line) model in comparison with the experimental data (symbols) {for various $V_p$ and $P_\mathrm{EC}$}. $V_p$ represents the gas puffing voltage which is proportional to the pre-filled gas amount.}
    \label{fig:exp_VS_PWSO0D_121224}
\end{figure}

According to the impurity radiation measured using the EUV spectrometer,
the main impurities in the plasma include carbon and tungsten,
which is sputtered from the divertor target plates. Previous experimental and modeling results demonstrate that carbon is a dominant impurity causing tungsten sputtering in L-mode plasmas on EAST~\cite{hxie2017,fding2019}.
Thus in the calculation of PWSO model, we consider the sputtering
of tungsten by carbon ions. The impurity radiation rate
$R_\text{coe,W}$ is estimated to be a constant value
$10^{-30}$ Wm$^{-3}$ based on the simulation results
using the FLYCHK code \cite{FLYCHK2008}.
We further assume that the perpendicular diffusion
coefficient $D_\perp$ of target impurities is
{$3$} m$^2$s$^{-1}$ \cite{FJCasson2014}, and one percent of the sputtered
atoms penetrate the main plasma \cite{AKirschner2017}, undergoing ionization at a
distance $\lambda=0.01$ m away from the target \cite{zywen2024}.
{A {maximal} projectile energy {$E$} of 740 keV, well above the thermal energy, is used as the upper limit in the integral in Eq.~$\left( \ref{eq:ITt} \right)$.}
For these experiments on EAST, the minor radius $a = 0.45$ m.
With the above EAST parameters, the PWSO 0D model predicts
that there are two density limit basins. The EAST experimental
results are located in the density-free regime, exceeding the Greenwald density limit (Fig.~\ref{fig:exp_VS_PWSO0D_121224}), and are in good agreement with the PWSO 0D model prediction. 
{The decreasing target region plasma temperature $T_t$ with discharge number (Fig.~\ref{fig:Tt_3D_022725}) also means that EAST is working in the virtuous process of the density freedom basin indicated by the PWSO model (section 4.1 of \cite{escande2022}).}

\subsection{PWSO 1D model and comparison}
The PWSO 1D model describes a more detailed evolution of the radiation power and the temperature towards the PWSO equilibrium
 \cite{escande2022}. 
    The impurity density and the plasma temperature evolution are determined from the
    following 1D transport equations
    
    \begin{equation}
      \begin{array}{rl}
        &\partial_{t} n_{i}-D \partial_{x}^{2} n_{i}=C_{i}\left[\partial_{x} T\left(r_{\text{LCFS }}, t-\tau_{\text {delay }}\right)+T_{\text {loss }}^{\prime}\right] \delta(x-a+\lambda)~,
      \end{array}
      \label{equ:nitran}
    \end{equation}
    \begin{equation}
      n \partial_{t} T-K \partial_{x}^{2} T=C_T T^{3/2}+P_{\text {add }}-n n_{i} \operatorname{Rad}(T)~,
      \label{equ:Ttran}
    \end{equation}
    where $n_i$ is the impurity density, $T$ the plasma temperature, $n$ the uniform plasma number density, {$C_{i}=-{f a K I\left(T_{\text{t}}\right)}/\left({(a-\lambda) T_{\text{t}}}\right)$} represents 
    the plasma-wall interaction, $K$ is a uniform diffusion coefficient, 
    {$C_{T}={E_{0}^{2}}/\left({\eta(T) T^{3 / 2}}\right) \simeq 6.510^{2} {E_{0}^{2}}/{Z}$}, 
    with $E_0$ the electric field corresponding
    to the loop voltage, $\eta(T)$ is the transverse Spitzer
    resistivity, and $P_{\text{add}}$ is the additional power
    density. For EAST experiments, the transport coefficient $\chi=K/n= 0.5~\mathrm{m}^2\mathrm{s}^{-1}$ is assumed. The same parameters as those used in the 0D model are applied to obtain the following results and the resistivity $\eta$ is considered to be a constant. Applying the following initial and boundary conditions
    \begin{equation}
  \begin{array}{ll}
      n_{i}(t=0)=0 ; & T(t=0)=T_{0} \\
    n_{i}(x=a)=0 ; & \left.\frac{\partial n_{i}}{\partial x}\right|_{x=0}=0 \\
    T(x=a)=T_{0} ; & \left.\frac{\partial T}{\partial x}\right|_{x=0}=0 
  \end{array}  
\end{equation}
at plasma center $x=0$ and edge $x=a$, the 1D model equations are numerically solved with various number density levels. The density limit is identified when the evolution of radiation power becomes unstable.
A relation between the density limit and the target region temperature has
been thus obtained and compared with the experimental data as shown in
Fig.~\ref{fig:exp_VS_PWSO0D_121224}.

The PWSO 1D model gives predictions similar to the PWSO 0D model. The EAST  experimental data are located in the density free basin,  as in the case of the 0D model. So, further altering the wall conditions, including increasing the pre-filled gas pressure and ECRH power, should result in a cleaner plasma and lower plasma temperature around divertor, and lead
to a higher density limit. 
\section{Discussion}
These EAST experiments show that a moderate increase of ECRH power and/or
pre-filled gas pressure at start-up enhances the density limit at the flat-top phase of discharge. 
This enhancement of the density limit is related to the lower
plasma temperature around divertor target and can be explained using the
plasma-wall self-organization model. Whereas the experimental results on
J-TEXT tokamak are found to locate in the density limit basin predicted by the PWSO theory \cite{jxliu2023},
the EAST experimental results are located in the density-free regime, which further
validate the PWSO theory and highlight the significance of high-Z materials
in the targets for accessing the enhanced density limit and the density-free regime {as expected in \cite{escande2022}} These experiments demonstrate and validate a practical scheme to raise the density limit, which can be extended and applied to other magnetic confinement fusion devices in future. 

In both EAST and J-TEXT experiments, some discharges with identical pre-filled gas pressure and ECRH power at start-up can find different plasma temperature around the target, which ultimately leads to different density limits.
These differences signify the effect of wall conditions on the high-density discharges.
How the varying wall conditions can affect the target region plasma temperature and the corresponding density limit may be used to further validate the PWSO model.

{Compared with W7-X \cite{Klinger2019,escande2022}, {which has a plasma volume $V = 30~\mathrm{m}^3$, a magnetic field $B= 3~\mathrm{T}$, and a minor radius $a = 0.53~\mathrm{m}$}, EAST has {$V=9 ~\mathrm{m}^3$, $B= 2.5$~T and $a=0.45~\mathrm{m}$}. The magnetic field $B$ and minor radius $a$ are close for the two machines. Then, it makes sense to compare the injected power per unit volume $P/V$ for EAST and W7-X experiments. One finds respectively {$P/V = 0.6~\mathrm{MW}/9~\mathrm{m}^3$} for EAST, and {$2~\mathrm{MW}/30~\mathrm{m}^3$} for W7-X, {which are exactly the same ratio $P/V=1/15~\mathrm{MW/m^{-3}}$}. 
In addition, in W7-X experiments, a density of $2 \times 10^{19} \mathrm{m}^{-3}$ (an order of magnitude higher than in tokamak breakdown) is reached with a single gas puff, right before injecting waves (Section 7 of \cite{escande2022}). In EAST, almost the same density is reached at the end of the ECRH pulse (Fig.~\ref{fig:discharges_differentPEC_121224}), which further increases to about $5\times10^{19} \mathrm{m}^{-3}$. These results demonstrate tokamaks can operate in a way similar to stellarator and reach likewise a higher density limit. The big difference is in the duration of the pulse: for W7-X, the start-up takes about 100 ms, while it is more than an order of magnitude longer in EAST to reach the same density with ECRH assisted start-up. This is reasonable because a tokamak needs a finite poloidal magnetic field to confine, while the confinement is available in a stellarator from the outset.}

Stellarator and tokamak experiments also find that the density limit increases with the heating power during flat top~\cite{Zanca2017,Fuchert_2018,Klinger2019,Wolf2019}. For EAST experiments in this paper, the ECRH power is only up to 600 kW. Together with the predictions of the PWSO theory (Fig.~\ref{fig:exp_VS_PWSO0D_121224}), it is expected that
higher ECRH power and pre-filled gas pressure in EAST experiments might lead to a detached plasma state
with a significantly higher density limit.

\section{Conclusion}\label{sec13}
In this article, we report on the experimental results on EAST tokamak
that have achieved the line-averaged electron density in the range of 1.3
to 1.65 Greenwald density limit, and our quantitative comparison
with the predictions of plasma-wall
self-organization theory that shows good agreement.
Increasing ECRH power and/or pre-filled gas pressure are confirmed to lead
to lower plasma temperature around divertor target and higher
density limit at flat top. The experiments are found to locate in the
density-free regime of PWSO model from both 0D and 1D predictions. These results demonstrate the
potential of a practical scheme for substantially increasing the density limit in tokamaks, which is also germane to the stellarator start-up. 

\section{{Methods}}
\subsection*{EAST}
The Experimental Advanced Superconducting
Tokamak (EAST) is a super-conducting tokamak with a major
radius of $R_0\leq1.9$~m, a minor radius $a\leq0.45$~m, plasma current $I_p\leq1$~MA, and toroidal field $B_t\leq3.5$~T~\cite{east_bnwan2022,xzgong2024}.
Two full tungsten divertors are located at the top and bottom of the toroidal vacuum chamber with full metal walls~\cite{lcao2015,gsxu2021,dmyao2015}. Various wall conditioning techniques, like lithiation~\cite{jshu2014} {and boronization, can be employed} in experiments. EAST is equipped with an auxiliary heating system with four gyrotrons~\cite{hdxu2016_EAST-ECRH,wyxu2024}. The working frequency of the ECRH system is chosen to be 105 or 140 GHz and the second harmonic extraordinary mode (X2) is used for electron heating and current drive~\cite{wwei2014_EAST-ECRH,hdxu2021_EAST-ECRH}. 
\subsection*{Techniques to obtain diagnostic data}
The line-averaged density is measured using three vertical chord HCN (hydrogen cyanide) interferometers and eleven horizontal chord polarimeter-interferometers~\cite{QXu2008,HQLiu2014}. A divertor triple probe system is embedded in the divertor target plates for the measurement of plasma density, temperature, particle, and heat
flux profiles along the target plate~\cite{TFMing2009,jcxu2016}.
The time and space resolved evolutions of radiation are measured using 64 channel absolute extreme ultraviolet (AXUV) photodiode arrays and the total radiation power is measured using a metal foil resistive bolometer~\cite{ymduan2015,wwen2024}.
The impurities in the core plasma are monitored with a flat-field extreme ultraviolet (EUV) spectrometer~\cite{lzhang2015}. {The effective ion charge $Z_\text{eff}$ is measured with the visible bremsstrahlung system \cite{yjchen2013}.}

\subsection*{Discharge parameters for density limit experiments}
{
Table \ref{tab1} summarizes {the} set of EAST discharges designed to explore the density limit {with} varying ECRH power and pre-filled D$_2$ amounts, along with the corresponding achieved density limits. }
\begin{table}[h]
\caption{
Summary of key parameters for the density limit discharges analyzed in this study. The first column lists the shot numbers. $P_\text{EC}$ in the second column indicates the applied ECRH power. The third column shows the pre-filled gas amount, which is positively correlated with, and controlled by, the applied pre-fill voltage. The fourth column presents the achieved line-averaged electron density, taken as a measure of the density limit.
}
\label{tab1}%
\begin{tabular}{@{}llll@{}}
\toprule
shot number & $P_\text{EC}$  & Pre-filled gas (D$_2$) amount & Density limit\\
(1430xx) & (kW)  & ($10^{20}$ molecules) & $(10^{19}$m$^{-3}$)\\
\midrule
64    & 0   & 0.66  & 4.8  \\
69    & 600   & 0.66  & 5.2  \\
73    & 600   & 1.0  & 5.3  \\
74    & 600   & 1.37  & 5.4  \\
75    & 600   & 1.73  & 5.4  \\
77    & 0   & 0.66  & 5.5  \\
79    & 600   & 0.66  & 5.6  \\
80    & 600   & 0.66  & 5.6  \\
\botrule
\end{tabular}
\end{table}

\subsection*{Calculation of PWSO 1D model}
The PWSO 1D model, which consists of a impurity particle transport equation (Eq.~\ref{equ:nitran}) and a heat transport equation (Eq.~\ref{equ:Ttran}), is solved using an implicit finite difference method. The resulting system of linear equations is solved using the Thomas algorithm (tridiagonal matrix algorithm).

\subsection*{Data availability}
Raw data were generated by the EAST team. Derived data that support the findings of this study are available from the corresponding author upon reasonable request.
\subsection*{Code availability}
The code used to produce the results in this study is available from the corresponding author upon reasonable request.

\backmatter

\bmhead{Acknowledgements}
This work is supported by the National MCF Energy R\&D Program of China under
Grant Nos.~2019YFE03050004 and 2022YFE03020004, the U.S. Department of Energy Grant
No.~DE-FG02-86ER53218, and the Hubei International Science and Technology Cooperation Project under Grant No.~2022EHB003. The computing work in this paper is supported by the Public Service Platform of High Performance Computing by
Network and Computing Center of HUST.




\bibliography{sn-bibliography}

\newpage

\newpage

\newpage

\newpage

\newpage
\end{document}